# Compositionally Graded SS316 to C300 Maraging Steel using Additive Manufacturing

A. Ben-Artzy[1], A. Reichardt[2], J-P. Borgonia[3], R.P. Dillon[3], B. McEnerney[3], A.A. Shapiro[3], P. Hosemann[2,4*]

1) Ben-Guriyon University, Material engineering Dep., POB 653, Beer-Sheva, 84105, Israel.
2) University of California at Berkeley, Department of Nuclear Engineering, CA, 94720, USA
3) Jet Propulsion Laboratory, California Institute of Technology, Pasadena, CA, USA
4) Lawrence Berkeley National Laboratory, Material Science Division, CA, 94720, USA

*corresponding author.*

Abstract

Joining of dissimilar metals is required for numerous applications in industries such as chemical, energy and automotive. It is challenging due to differences in melting point, density, and thermal expansion of the metals being joined. Common welding techniques involve limiting melting and solidification to a narrow area leading to high thermal stresses and potentially brittle intermetallic phases. Furthermore, the geometric complexity of these welded joints can be rather limited. Additive Manufacturing (AM) presents new techniques for joining of dissimilar metals. One of the emerging methods is the building of functionally graded parts using Directed Energy Deposition (DED) to spatially vary composition. In this paper, a SS316L and C300 maraging steel couple were joined by DED and heat treated. 13 discrete composition layers were selected using metallurgical considerations, in order to ensure a smooth transition in properties and microstructure. The mechanical properties of the as-built joints were found to be similar to the SS part and no intermetallic phases were found in the interface.

1. Introduction and Background

Additive Manufacturing (AM) allows for fabrication of complex geometries without elaborate tooling to produce a near net shaped component. Numerous innovative AM techniques are available today, among them the Directed Energy Deposition (DED) which is known for its high deposition rate of metal powders. The build rate can be as high as 0.5kg/hr, but faster build rates are still desired for some industries [1-3]. One of the advantages of DED compared to Laser Powder Bed Fusion (L-PBF) is the ability to change the powder feed when multiple powder hoppers are deployed. This allows the chemical composition to be continuously changed without the need to open the machine or interrupt the process. This feature presents the opportunity to manufacture compositionally graded parts, compositionally clad parts, or parts with homogenous regions connected by a region of gradient composition (as in the current research) [4-7].

Maraging steels are high strength alloys that obtain their strength from intermetallic nanoprecipitates, while 316L stainless steel (SS) offers good toughness at cryogenic temperatures due to its austenitic structure. The C300 maraging steel is classified as a low carbon precipitation hardening martensitic steel. As an ultrahigh strength cobalt-containing steel alloy, C300 was design for wear resistance in rough environments [2,7,8] while austenitic stainless steels are paramagnetic and corrosion resistant. The combination of both allows the design of components requiring localized wear resistance (like bearings) and localized paramagnetic behavior and corrosion resistance. Some of the difficulties joining these materials are due to the major differences in the mechanical and physical properties of the two metals, as can be seen in Table 1. The coefficient of thermal expansion and the heat capacity are almost double for SS316 compared to C300, and the mechanical properties are also significantly different. In addition, the post fabrication heat treatment is significantly different between the two materials. The strengthening mechanism of the C300 maraging steel is based on precipitation hardening mechanism of molybdenum-based precipitates, such as $Ni_3Mo$, $Fe_7Mo_6$, $Fe_2Mo$, and σ-FeMo [8-11], which form after an aging heat treatment. Unlike carbon steels, the C300 maraging steel is relatively soft prior to aging, and the precipitation hardening mechanism is used to set the strength and hardness of the material allowing for nearly no shape change during the final heat treatment. The aging of C300 maraging steel is 482°C for 3-8hr. One of the only common properties for both steels is the very low carbon content.

**Table 1: Physical properties of C300 maraging steel and SS316L stainless steel (SS) [3].**

| Property | C300 maraging steel | SS316L |
|---|---|---|
| Crystal structure | BCC Fe-Ni martensite + intermetallic precipitates ($Ni_3Mo$ & $Ni_3Ti$) | FCC austenitic |
| Major alloying elements | Ni, Co, Mo, Ti | Cr, Ni, Mo, Mn |
| Carbon content [wt.%] | < 0.03 | < 0.03 |
| Thermal expansion [μm/m-°C] | 8.64 | 16.0 |
| Specific heat capacity [J/g-°C] | 0.335 | 0.500 |
| Thermal conductivity [W/m-K] | 19.6 | 16.2 |
| Yield strength [MPa] | 827 (annealed)    1930 (aged) | 290 |
| Tensile strength [MPa] | 1030 (annealed) | 558 |
| Strengthening mechanism | Precipitate strengthened | Cold working |
| Standard heat treatments | Annealing: 815 °C, 1h | Annealing: 1040 – 1175 °C |
| Magnetism | Ferromagnetic | Paramagnetic |
| Liquidus [°C] | 1454 | 1440 |
| Solidus [°C] | 1427 | 1390 |
| Density [g/cm3] | 8.00 | 7.99 |

The nominal chemical composition of the two steels is listed in Table 2. It can be seen that the significant chemical composition difference lies in the cobalt and titanium content on the C300 side and high chromium content on the SS316L side. Nickel and manganese are also missing or reduced on the C300 maraging steel side, and the Molybdenum content on the C300 side is twice that of the 316L.

Table 2: Nominal chemical composition of C300 maraging steel and 316L steels [9,10,23] in wt%. The respective Ni and Cr equivalent range is also provided for guidance in figure 1 (Schaeffler diagram)

|      | Fe  | Cr   | Ni        | Mo      | Co      | Mn     | Ti      | Ni equ. | Cr equ. |
|------|-----|------|-----------|---------|---------|--------|---------|---------|---------|
| C300 | Bal | <0.5 | 18.0-19.0 | 4.6-5.2 | 8.5-9.5 | 01 Max | 0.5-0.8 | 19      | 5       |
| 316L | Bal | 16-18| 10.0-14.0 | 2.0-3.0 | ---     | 2.0 Max| ---     | 12      | 19      |

The Schaeffler diagram, Figure 1 [22], provides guidance for the gradient path selection from C300 to SS316L. It can be seen that C300 will have a mix of austenitic/martensitic structure transitioning to full austenite at composition of ~65% wt. SS316L. The nominal composition of SS316L would have a mostly austenitic structure with approximately 5% ferritic content. It is worth noting that the tolerances for most alloys are rather large and the exact position in the Schaeffler diagram may vary as outlined in the shaded region in Figure 1 and tend to be on the lower end of the nickel equivalent due to nickel cost. The nickel equivalent to chromium equivalent ratio (on the SS side) can vary between 0.553 and 0.815 depending on the specific composition of the material, but still fulfilling the requirements for standard 316L stainless steel.

Laser deposition allows this transition in composition be done in steps over relatively long distances. Commercial laser deposition systems are capable of achieving composition change as small as 1% with each layer [21,26]. The final result can be a smooth gradient in microstructure and properties.

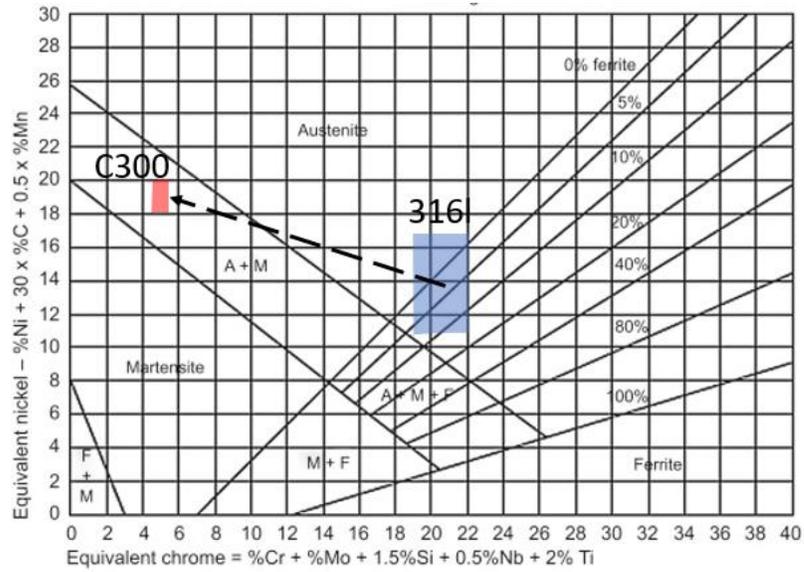

**Figure 1: Linear gradient path between C300 maraging steel and SS316L on the Schaeffler graph**

The formation of intermetallic phases in any dissimilar metal joint maybe one of the biggest challenges encountered. Few binary systems are characterized by complete solubility over wide range of compositions and temperatures. Furthermore, one should consider not only the major ternary alloying elements, but also minor alloying elements and impurities which can stabilize or produce deleterious phases as well. The relevant binary diagrams are provided in Fig. 2. By contrast the binary systems Fe-Ti, Fe-V, Ni-Ti, Ni-V and Cr-Ti all contain a number of thermodynamically stable intermetallics that are known to be hard and brittle. The Fe-Ti system showing FeTi and $Fe_2Ti$ contain a large field corresponding to a hard, brittle phase. The Ni-Ti system with $TiNi_3$, TiNi, and $Ti_2Ni$ contains intermetallic compounds in addition to the sigma phase, and the Cr-Ti system is characterized with a brittle $TiCr_2$ intermetallic (see Figure 3).

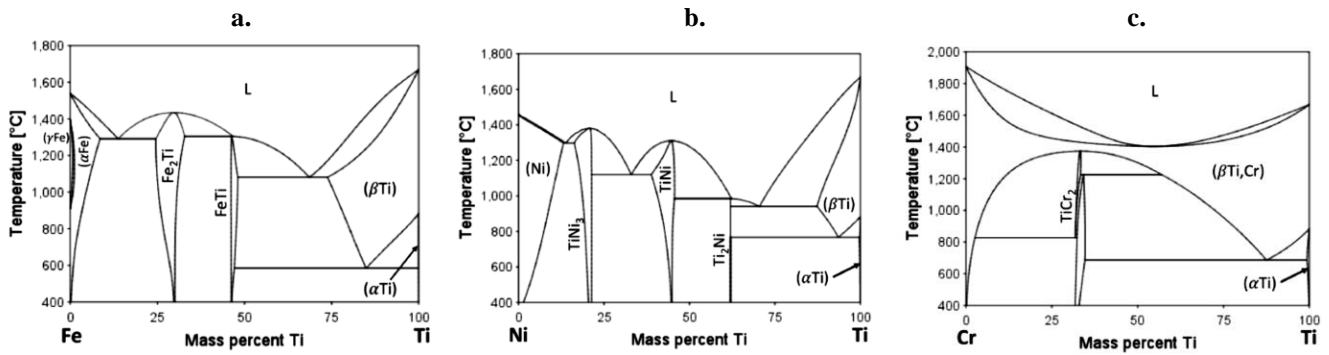

**Figure 2: Binary phase diagrams of the (a) Fe-Ti system, (b) Ni-Ti and (c) Cr-Ti. All diagrams were generated with Thermo-Calc using the TCBIN binary database.**

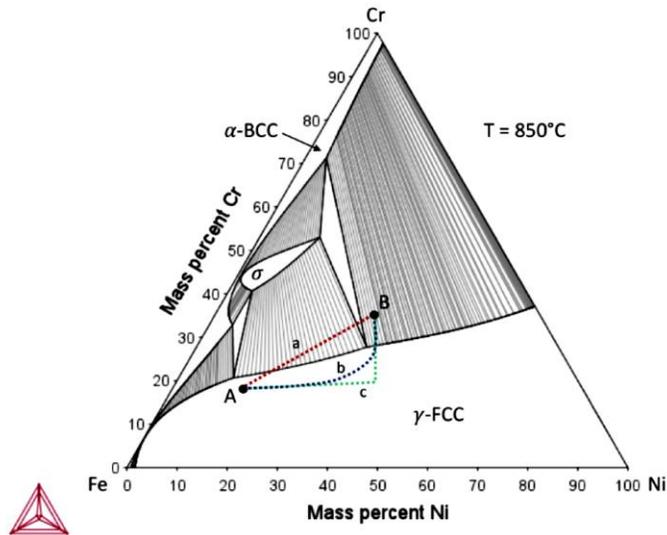

Figure 3: Ternary phase diagram of Fe-Cr-Ni used to choose the gradient path between "A"-316L Austenitic stainless steel and "B"-C300 Martensitic Maraging steel. Diagram was generated with Thermo-Calc TCFE7 database.

One advantage of processing multiple materials with DED is that layer-by-layer deposition allows for any composition path to be followed. The manufacturing path chosen is based on thermodynamic modeling in order to identify a composition space such that regions of insolubility and intermetallic formation are strategically avoided. Figure 3 shows an example of joining two Fe-Ni-Cr alloys plotted on an isothermal ternary section. Alloy A is the 316L austenitic stainless steel and B is the C300 maraging steel. Under equilibrium conditions, alloy A is expected to be fully austenitic, while alloy B is a dual phase austenitic-ferritic structure. It is obvious that following the "a" path leads to intersecting two and three -phase fields that include the hard, brittle σ-phase. Instead, a more favorable albeit non-linear path "b" circumvents compositions at which σ-phase formation is likely. In path "c" the phase evolution along the gradient path consists only of ductile solid solution of γ-austenitic and α-ferrite, rather than brittle intermetallic phases. In choosing a ternary gradient path (at 850°C in Figure 3), it is also necessary to consider the system over a range of temperatures.

In order to achieve maximum strength for the C300 maraging steel it needs to undergo a solution heat treatment at 815°C for 1hr followed by air cooling. The cooling after the solution heat treatment stabilizes a martensitic structure. Due to the high amount of nickel and low carbon content, the martensite structure is not expected to be distorted or to have a BCT structured, nevertheless is a rather soft BCC martensite. An aging heat treatment is then done at 482°C for 3-10 hours to achieve the final strength and hardness. Since 316L austenitic stainless steel is a single phase steel, the solution heat treatment seems to not alter an AM produced 316L. Holding the gradient joined couple for a long time in the solution temperature might cause grain growth in the SS side. The only ways to increase the strength of SS316 side of the joint is by cold work. A surface treatment such as nitriding can increase the surface hardness but neither strengthening nor hardening the SS316 was desired in this work. The conclusion is that the joint strength will be limited by the 316L side, assuming that the interface strength is not weaker than the stainless steel.

This work utilized AM to create a durable transition joint between two structural alloys. The gradient path was carefully selected in order to prevent the formation of intermetallic brittle phases and provide a transition between SS316L Austenitic Stainless steel and 300 Maraging steel via Laser deposition of 13 discrete composition layers Post build and heat treated mechanical properties and microstructural characterization were carried out.

## 2.    **Experimental**

### 2.1.    *Manufacturing of the functionally graded composition of the joint*

A compositional gradient between the C300 maraging steel and SS316L was deposited using RPM Innovations, Inc. 222 DED system. A summary of the system specifications is provided in Table 3. The RPM 222 DED, is equipped with a 2'×2'×2' (XYZ) work envelope and 4 axis motion (XYZ and R). Deposition was performed with an IPG fiber laser operating at a wavelength of 1060 – 1080 nm and power up to 2 kW. The process parameters for each alloy (laser power, scan speed, beam spot size, and powder mass flow rate) were based on prior operating experience. The laser scan speed was maintained at a fixed value of 30 in/min (1.27 cm/s), and the layer thickness was 0.508 mm. The laser power settings are shown in Table 4. In practice, the measured laser power differed from the nominal set point by as much as a few percent, and was considered to be the actual value. Hatch width was typically 0.889 mm. The ideal powder mass flow rate was determined experimentally for the combination of listed parameters (laser power, layer thickness, and hatch width). All samples were fabricated in an argon atmosphere glove box with oxygen level controlled to <10 ppm. The build plates or substrates were comprised of 0.25in. (0.63cm) thick plates of SS alloy.

To achieve compositional grading, the RPM DED systems used were equipped with two powder hoppers and feeders allowing two distinct alloy powders to be individually supplied, mixed, and simultaneously deposited. For each composition step, the powders were mixed *in situ* prior to exiting the build head nozzles during deposition. This was done by varying the flow rate in the powder feeders to achieve the desired volume fraction of each powder type. The number of layers per composition step and the magnitude of the steps, were incorporated in a predetermined build plan that was carried out during deposition. It is important to note that the use of additional powder feeders would enable a more complex gradient path that could be followed. The build started with 100 layers of C300 maraging steel, and then gradually transitioned to SS316L through a smoothly transitioning gradient region. The gradient region consisted of 13 layers, with a composition change of 7.134% by volume each layer according to Table 5. The layer thickness was 0.508 mm, making the gradient region ~6.6 mm wide (see Figure 4a).

The scan pattern consisted of a common outlining (aka contour) plus bi-directional raster sequence (aka hatch). With this method, the build head begins by outlining the contour of the layer, and then fills in the remaining area by rastering back and forth or hatching. To ensure even building, the starting point of the sequence and the hatch angle are alternated in each layer. Dwell time between the layers was marginal.

The feedstock material consisted of standard gas atomized or PREP powders compatible with laser deposition processing. The size distribution was controlled by the powder manufacturer through sieving with mesh sizes ranging from –60 to +325 depending on the powder type, corresponding to particle sizes ranging from 45 $\mu$m to 250 $\mu$m. All powders were received with chemical analysis showing the content of significant elements. The chemical composition of the powders used in this study are provided in Table 6. In comparison to Table 2 all elements are in the acceptable composition range except the Ti in C300 which exceeds the maximum of 0.8 allowed by the ASTM standard. During deposition, the mixed powder was blown into the melt pool through a set of 4 nozzles located on the build head with the aid of argon gas flow.

**Table 3: RPM DED 222 Innovations Inc. laser Deposition system use in this study.**

| | |
|---|---|
| **Work envelope** | 2ft × 2ft × 2ft (0.61m × 0.61m × 0.61m) |
| **Motion control** | 4 axes (CNC X/Y/Z + rotate table) |
| **Laser** | IPG fiber laser |

| Maximum laser power | 2 kW |
|---|---|
| Laser wavelength | 1060 – 1080 nm |
| Atmosphere | Argon (<10 ppm $O_2$) |
| Scan pattern | Outer contour plus bi-directional raster with alternating starting point and angle |
| Powder delivery system | Powder delivery system |
| Deposition head | Water cooled 25-degree 4 nozzle |

**Table 4: Process parameters for maraging to stainless steel gradients.**

| Substrate | Powders | Layer thickness (mm) | Laser Power (W) | Hatch width (mm) | Spot size (mm) | Scan speed (cm/min.) |
|---|---|---|---|---|---|---|
| stainless | 316L stainless steel | 0.508 | 874 | 0.889 | 1.415 | 76.2 |

**Table 5: The composition changes from maraging to stainless steel gradients.**

| Layer | Vol. % C300 | Vol. % SS316L | g/min. C300 | g/min. SS316L |
|---|---|---|---|---|
| 0-99 | 100 | 0.000 | 11.928 | 0.000 |
| 100 | 92.857 | 7.143 | 11.076 | 0.388 |
| 101 | 85.714 | 14.286 | 10.224 | 0.810 |
| 102 | 78.571 | 21.429 | 9.372 | 1.239 |
| 103 | 71.429 | 28.571 | 8.520 | 1.677 |
| 104 | 64.286 | 35.714 | 7.668 | 2.122 |
| 105 | 57.143 | 42.857 | 6.816 | 2.577 |
| 106 | 50.000 | 50.000 | 5.964 | 3.040 |
| 107 | 42.857 | 57.143 | 5.112 | 3.514 |
| 108 | 35.714 | 64.286 | 4.260 | 3.997 |
| 109 | 28.571 | 71.429 | 3.408 | 4.493 |
| 110 | 21.429 | 78.571 | 2.556 | 5.000 |
| 111 | 14.286 | 85.714 | 1.704 | 5.520 |
| 112 | 7.143 | 92.857 | 0.852 | 6.054 |
| 113-213 | 0.000 | 100 | 0.000 | 6.604 |

**Table 6: Chemical composition of C300 maraging and 316L stainless steels.**

| Powder Name | Chemical Analysis – Wt. % | | | | | | | | | | | | |
|---|---|---|---|---|---|---|---|---|---|---|---|---|---|
| | Fe | Cr | Ni | Mn | Co | Ti | Mo | Si | S | P | C | O | N |
| 316L Stainless Steel | bal | 16.3 | 10.3 | 1.31 | – | – | 2.09 | 0.49 | 0.006 | 0.026 | 0.026 | – | – |
| C300 Maraging Steel | bal | 0.02 | 18.0 | 0.02 | 9.3 | 1.1 | 4.7 | 0.02 | 0.01 | <0.01 | 0.02 | 0.02 | <0.01 |

Following deposition, the cylinders were removed from the baseplate, further sectioned, and cut vertically with wire EDM such that the cross-sectional surface of the gradient could be polished for characterization (Figure 4a). Following the EDM sectioning, it was observed that the cut as-built halves were slightly bowed, indicating relaxation of residual stresses.

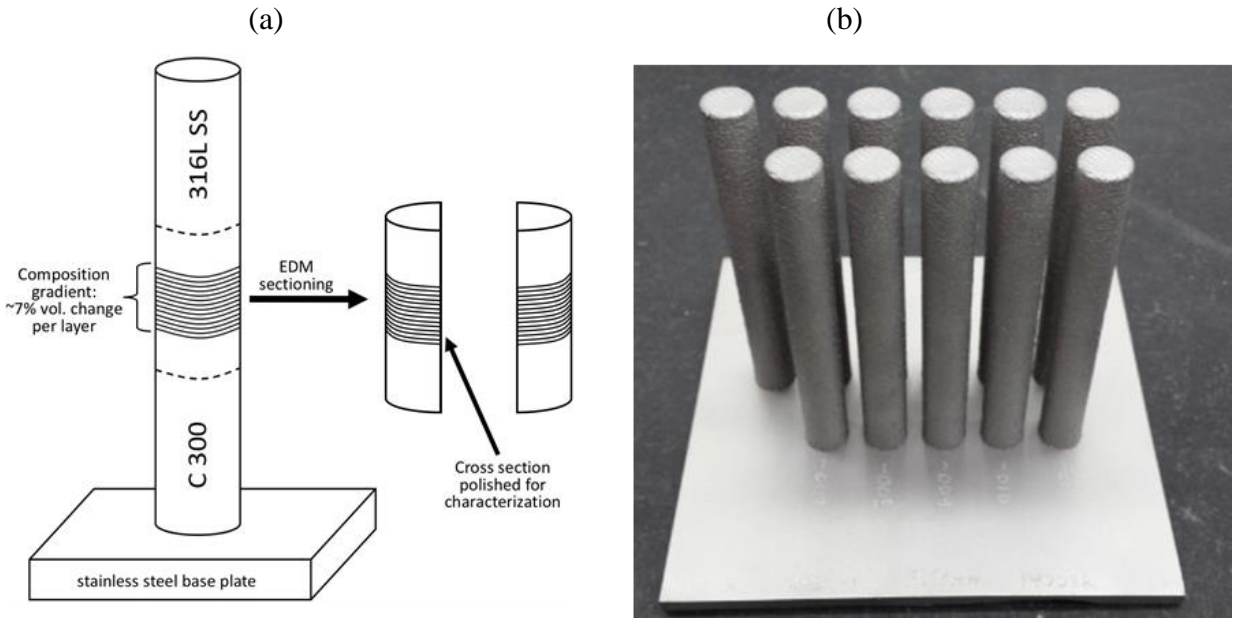

**Figure 4: (a) Schematic drawing of C300 maraging steel to SS316L sample fabricated and characterized and (b) image of DED fabricated samples attached to the baseplate.**

A set of as-built samples were then solution heated at 815 °C to fully austenize followed by air cooling to form a martensitic structure in the maraging steel. Subsequently, the samples were aged at 482°C for 6hr to allow for precipitation hardening of the maraging steel such that the alloy's maximum hardness values can be achieved, with negligible effect on the SS316L due to the low carbon content.

Following fabrication and heat treatment, the gradient samples were sectioned and prepared for microstructural analysis by grinding with 320 – 4000 grit SiC grinding papers, followed by polishing with 3 μm, 1 μm, and 0.1 μm diamond solutions, and a final polish of 0.02 μm non-crystallizing colloidal silica suspension. For microstructural characterization, the gradients were etched by swabbing with Kalling's No. 2 (5g $CuCl_2$ + 100mL HCl + 100mL ethanol) on the maraging steel half, and Glyceregia (15mL HCl + 10mL glycerol + 5mL $HNO_3$) on the stainless-steel half. Comprehensive microstructural mapping of the as-built and age-hardened gradients was performed with electron backscatter diffraction (EBSD), as presented in the following sections. For this study, acquisition of EBSD data was performed with an Oxford NordlysNano detector using the accompanying AZtecHKL software. Specimens were adhered with superglue or crystal bond to standard SEM stubs, and mounted in the SEM chamber on a 70-degree holder facing the EBSD detector. The SEM accelerating voltage was maintained at 20kV, and the beam current ranged from 8 to 16 nA. In general, higher beam currents were successful in improving the EBSD signal. In most cases energy dispersive spectroscopy (EDS) maps were acquired simultaneously. This required that the beam current be balanced to provide a sufficient signal for EBSD while not increasing the dead time for the EDS detector beyond a reasonable level. SEM and EDS chemical composition analysis were also used in the fractography evaluation of the sample's fracture surface following tensile testing.

*2.2. Mechanical properties evaluation*

Vickers hardness testing was performed along the composition gradients using a fully automatic Leco AMH55 hardness testing system. The tool was routinely calibrated with standard samples.

Since the sample is only 14mm in diameter, a sub-sized tensile test samples needed to be used. A SSJ-2 16mm long sample geometry was selected (See Figure 5a) for both the as-built material and heat treated samples. Furthermore, these samples are commonly used in the nuclear materials field and therefore the data can be used for nuclear applications. The sample thickness was 1mm in thickness for the SSJ-2 samples in order to comply with the ASTM standard [12-20]. All samples were ground and polished up to 600 mesh for measurement consistency.

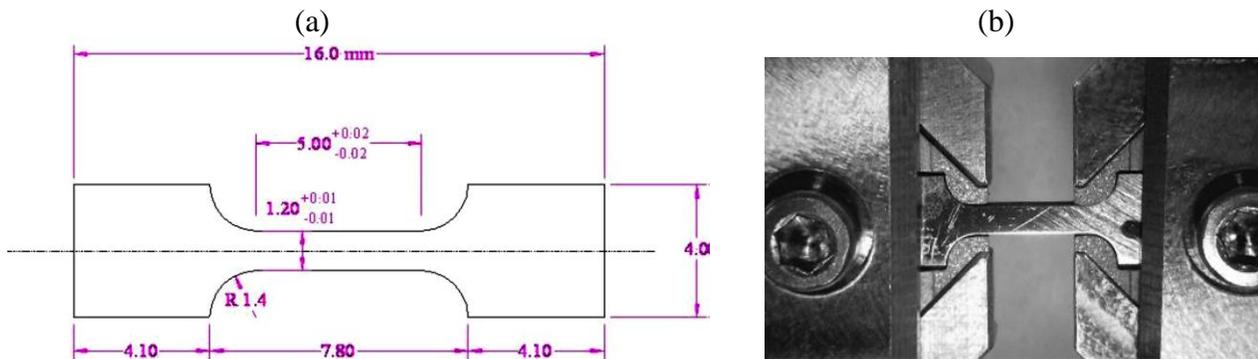

**Figure 5: (a) 1 mm thick SSJ-2 miniature tensile test sample schematic and (b) EDM wire cut SSJ-2 tensile test sample placed in a miniature tensile test machine.**

The miniature tensile test samples were tested in a Kammrath & Weiss miniature tensile test machine (Figure 5b) using 5KN load cell and MDS software control system. The tensile test was performed at a speed of 15μm/sec using beam movement as a movement control.

**3. Results**

The chemical composition of the interface was determined using EDS (Figures 6). The measured weight percentages of Co and Cr along the length of the sample are plotted, showing a near-linear composition slope along the gradient despite the stair-step deposition strategy. The interfacial region of compositional change is approximately 7-9 mm wide and is indicated in the figure by the two vertical lines. We do want to note that the starting point of the scans (X-axis) between Figure 6 and Figure 10 is different due to the long build length of the component. But the relevant section is marked as transition region in both figures and we aim to show the entire length collected. The linearity of the compositional gradient is explained by the fact that underlying layers are always remelted and mixed with the next deposited layer.

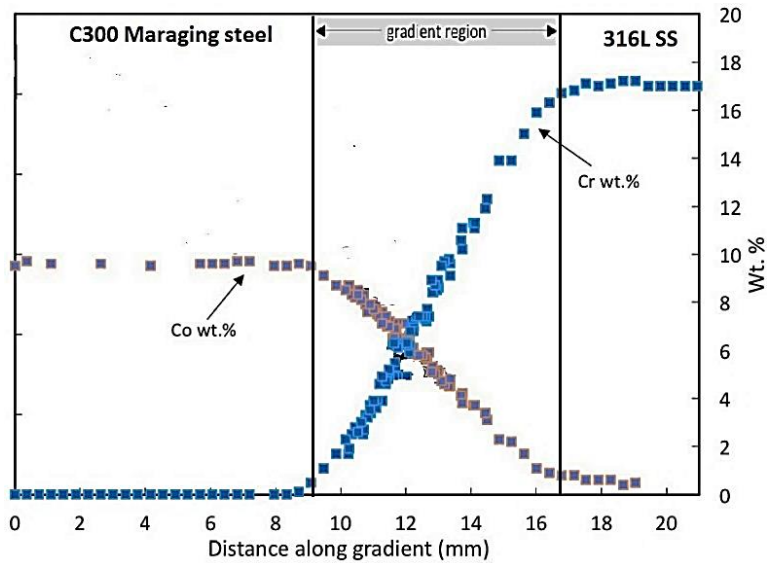

**Figure 6: Chemical composition across the interface [3]**

SEM of the etched sample was performed at different regions of the build. It was found that the microstructure shows solidification cells. The average cell spacing was estimated from the SEM micrograph in Figure 7 to be ~ 8 $\mu$ m, giving an estimated cooling rate of 2.15 $\times$ $10^3$ °C/s [2]. This is within the range expected for laser deposition [26-28]. This cellular structure is common for AM builds as the laser beam is moving in a perpendicular direction to the image surface. 2.15 $\times$ $10^3$ °C/s is consider as a rapid cooling rate therefore all rapid solidification known phenomena are expected. Further small pores between the cells can be found as shown in figure 7.

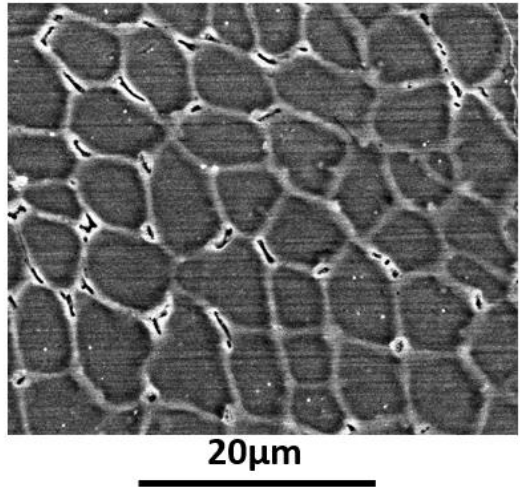

**Figure 7: SEM micrograph of solidification cells in etched as-built gradient at 100% SS316L.**

Etching the as-built and age-hardened gradients revealed a fine substructure throughout the composition gradient consistent with mixed cellular and dendritic solidification modes. The optical micrographs in Figure 8 shows an example of this substructure for both the 100% and 16% wt. SS316L regions. The solidification grain boundaries are visible, each of which formed via the solidification of fine cells or dendrites with shared orientation. Also visible are the melt pool boundaries of the deposited material, which appear as discontinuities in the cellular or dendritic substructure. The scale of the solidification cells is related to the cooling rate.

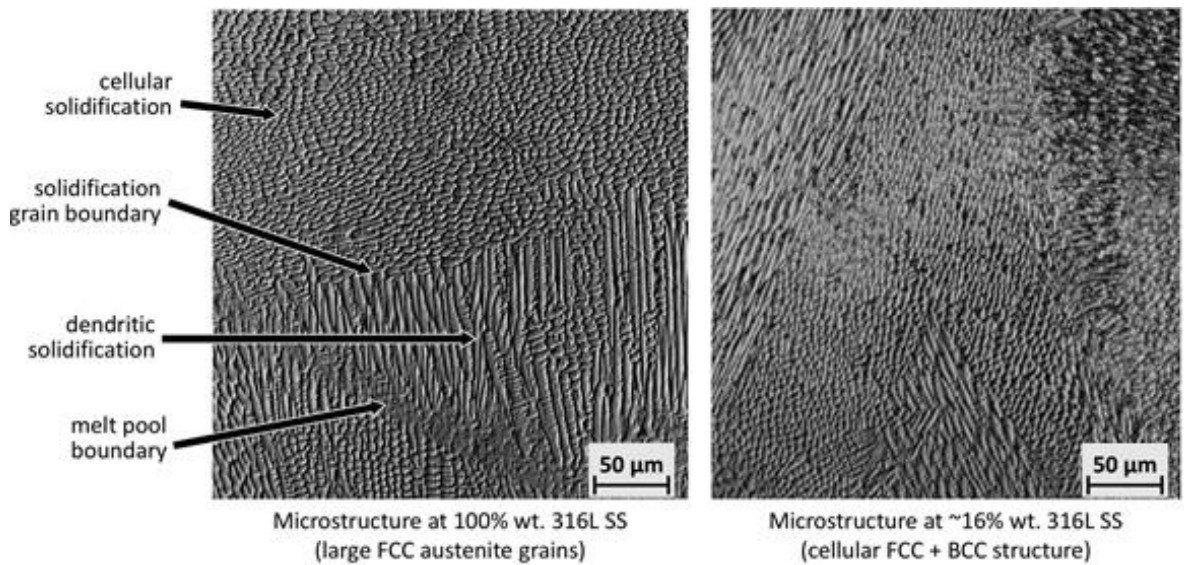

**Figure 8: Optical micrographs of solidification substructure in the etched as-built gradient at (a) 100% wt. SS316L and (b) ~16% wt. SS316L.**

EBSD mapping of microstructural evolution along the composition gradients in the as-built and age-hardened conditions is given in Figure 9. The complete map was reproduced from parts and indexing was performed for Fe-FCC and Fe-BCC only. No evidence of other secondary phases was detected. The microstructure along the gradient can be divided into three distinct regions. From the pure maraging steel to about 20% wt. SS316L, the material consists of fine irregularly shaped grains of BCC, with small amounts of retained austenite forming between them. Although EBSD cannot distinguish between ferrite and martensite, the BCC is presumed to be martensite as expected for C300 maraging steel. The range of 20% and 54% wt. SS316L then corresponds to a fine dual-phase region, consisting of BCC cells with FCC located in the cell boundaries. The solidification structure influence on SS316L microstructure transitions can be seen in large austenite grains 100-250 $\mu$m in diameter and growing across the interface layer. In the 100% SS316L layers, the grains continue to grow larger and more irregular. Minimal differences are seen between the as-built and age-hardened gradients, and the composition thresholds for major microstructural transitions remain unchanged after aging. The only apparent difference is a slightly increased percentage of austenite in the aged C300, which may be attributed to austenite reversion. In figure 9, the EBSD phase maps (upper scan) and Euler color grain orientation maps (lower scan) allow comparison between various regions along the gradients for as-built (as-deposited) and age-hardened conditions. The percentages on the left refer to the range of SS316L weight percentages that corresponds to the depicted microstructure. Stitched together EBSD maps showing the evolution of microstructure along the gradient from ~28% to 62% wt. SS316L, including phase and IPF X (orientation with respect to build direction).

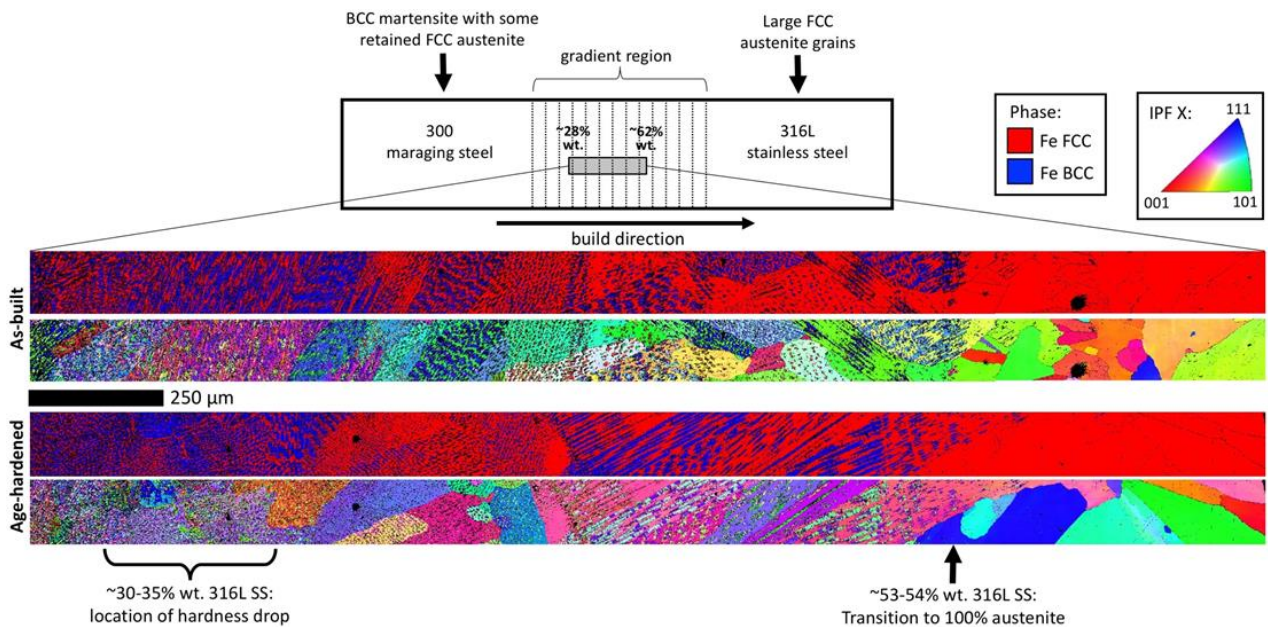

**Figure 9: EBSD maps showing the evolution of microstructure along the gradient from ~28% to 62% wt. SS316L, including phase and IPF X (orientation with respect to build direction) [3].**

Figure 10 shows the hardness results plotted as a function of distance along the samples, across the interface. Also indicated are commercially reported hardness values for conventionally processed SS316L (148 HV), annealed C300 (302 HV), and age-hardened C-300 (544 HV). The composition scale (the gray area) is only an approximate illustration since it was established using EDS. The hardness profiles are consistent with the previous set of data, and show that the hardness begins declining gradually with the addition of SS, drops abruptly through the range of 30-35% wt. The hardness change toward the interface can be related to solution hardening while the hardness drop is associated to the lack of age hardening elements in SS316L. The hardness then remains nearly constant through the rest of the composition gradient. The Vickers hardness results presents a higher resolution indent array made along as-built and age hardened gradients. Results are plotted in the as built, aged, annealed (solution heat treated), and solution and aged condition. The solution and aged condition have been chosen as the prime heat treatment to obtain fully aged properties of C300 with little to no impact on the 316L side.

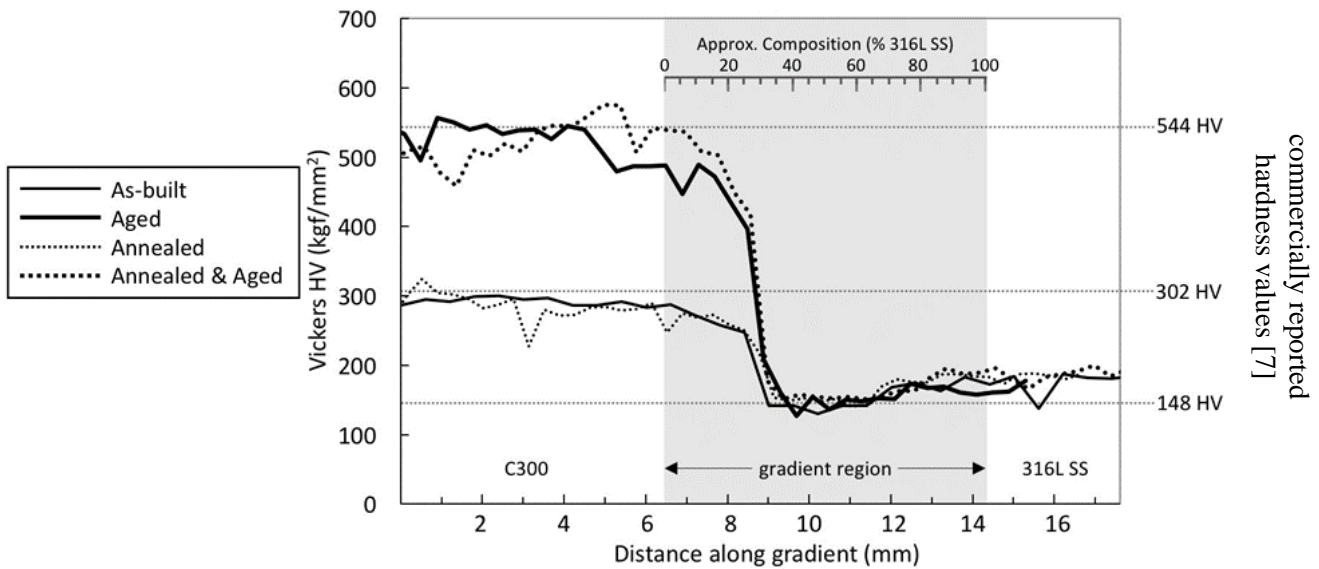

**Figure 10: Vickers hardness along C300 to SS316L gradients, of various heat treatment conditions (as-built, aged, annealed, and both annealed and aged). The horizontal dotted lines indicate commercially reported hardness values for SS316L (148 HV), annealed C300 (302 HV) and age-hardened C-300 (544 HV).**

The SS316L does not benefit from the solution heat treatment of the gradient as the solution heat treatment at this temperature does not result in any noticeable microstructural differences. The most notable feature in the hardness profiles is of course a steep drop in hardness that occurs at a precise composition threshold in all heat treatment conditions, corresponding to roughly 30-35% wt. SS316L, as shown in Figure 10. Its presence both before and after aging suggests that a major microstructural change occurs at this threshold after deposition, which subsequently affects hardenability during aging. However, given the layer thickness of ~500 $\mu$m, the spatial resolution of these Vickers measurements (~400 $\mu$m) is insufficient to determine whether the hardness drop occurs suddenly at a layer interface, or more gradually through the thickness of a layer or two. Following indentation, EDS was used to measure the composition at each indent, with the % wt. of SS316L approximated from the Co and Cr contents. This is an important step since each layer interface is not actually linear, but rather takes the form of a scalloped edge created by the laser melt pool tracks, as can be seen in Figure 11 where there is a clear change in etching response corresponding to the decrease in hardness along the composition gradient. Fig 11 shows the microstructural region where, due to composition change, age hardening is not effective that corresponds to the hardness decrease in Fig 10. It appears that the harder region also etches differently as can be seen in this image taken after etching. There appears to be no correlation to the phase fractions observed but correlates to the mechanical properties. This is likely due to the fact that Co and Ti in the maraging steel can have a strong effect on precipitation hardening and etching (or corrosion) behavior, but a small effect on the bulk phase fractions.

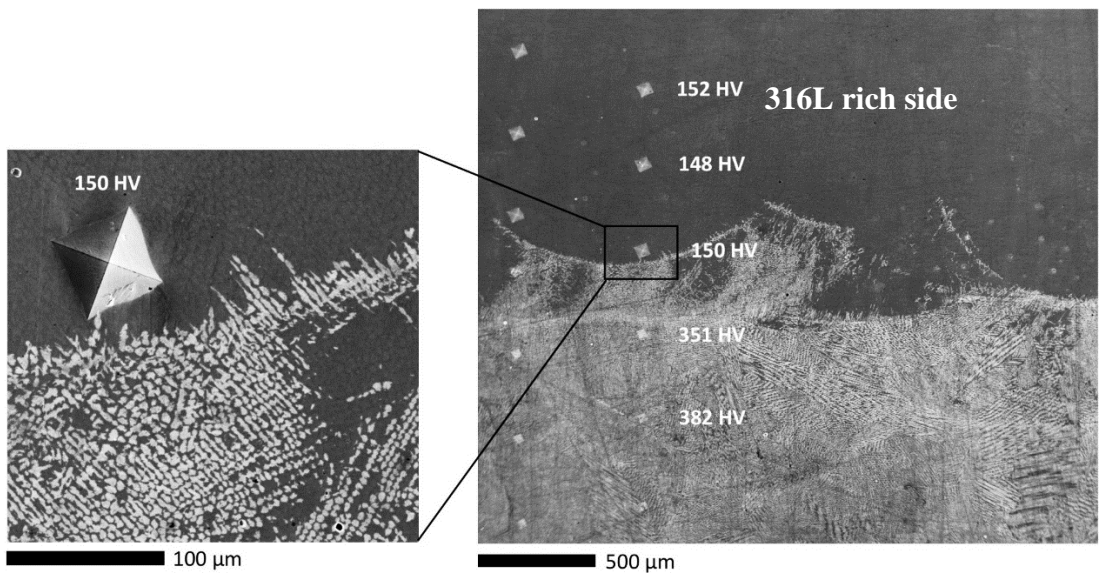

**Figure11: Vickers indent array on the age-hardened gradient.**

The results for the hardness testing of the as-built base alloys appears to be a slight increase in hardness at higher SS316L contents. To confirm the hardness numbers and reduce uncertainty 50 indents were made in either bulk alloy. The SS316L was found to have an average hardness of 200 HV, while the C300 had an average hardness of 300 HV.

Tensile tests were performed for both the as built and aged condition. Samples were wire EDM cut out of the interface zone and the base metal away from the interface in order to use as a reference test group. The miniature tensile test samples were polished prior to testing in order to remove any defects and scratches from the surface and in the gage region. Figure 12 shows the stress-strain curve of the interface area as-built (long-dashed, line) and after ageing (Black, solid, line) in comparison to the stainless-steel base metal (short-dashed line). It can be seen that the as built interface has the same YS and UTS of the SS while the elongation to fracture is reduced by half. The aged interface tensile strength is some 10% less than the stainless-steel while it's elongation to fracture is much of the same as the as-built interface.

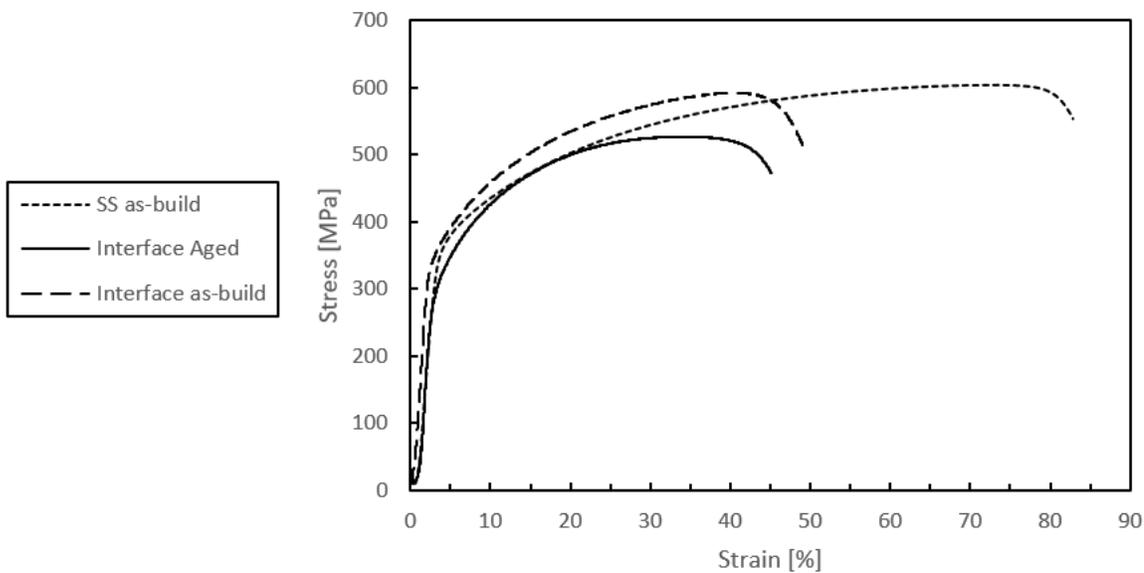

**Figure 12: Stress-strain curve of the interface zone as-built vs. aged (heat treated) in comparison to the stainless-steel side as-built.**

The interface of the aged tensile test sample can be seen in Figure 13. Indications of the melt pools can be seen clearly on both sides of the sample and the deformation zone can be distinguished due to its response to the etchant. The maraging steel on the left side is fully aged (HV567) while the SS hardness is almost not affected. The chemical composition, as evaluated using EDS, is presented on the top side of the picture, revealing the interface. Some of the deformation was experienced on the SS side while in the maraging steel deformation took place at the area of low Co, Mo and Ti, components that increase strength in C300.

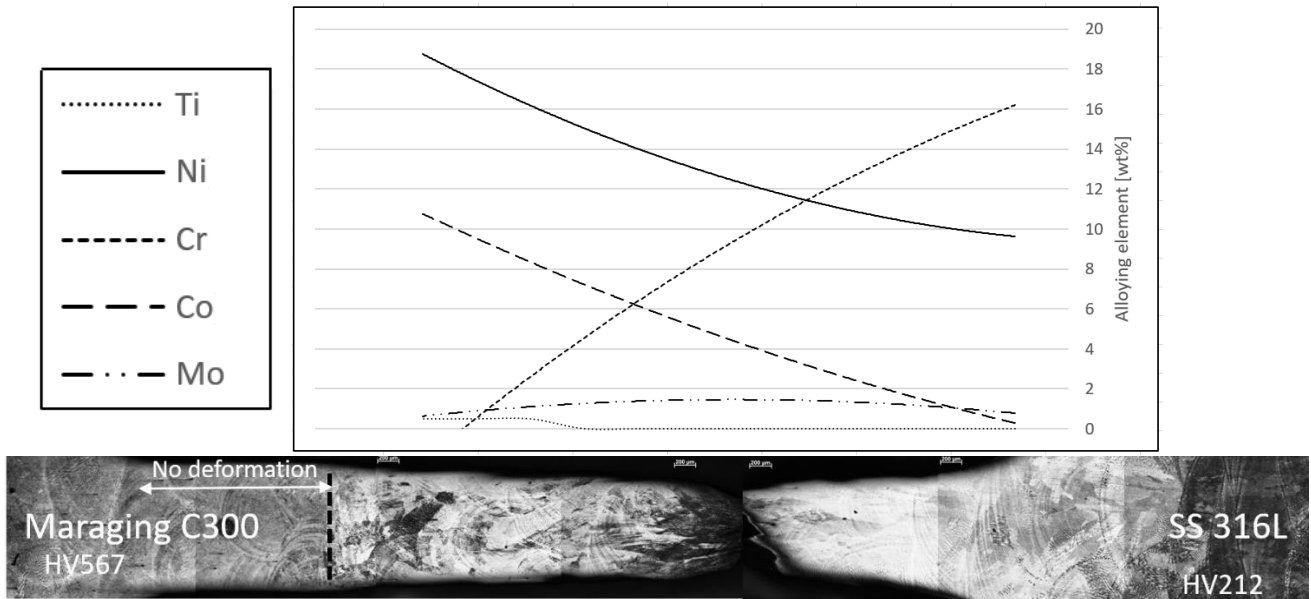

**Figure 13: Interface of aged tensile test samples.**

Fractography view of the tensile test surfaces is presented in Figure 14a, 14b aged, and in comparison, to the as built stainless steel 14c. It can be noticed that all fractures are ductile with the typical fine dimple structures. The dimples of the as-built interface sample are maybe slightly coarser (2.5μm in diameter) then the aged ones (2.2μm in diameter). The 79% elongation to fracture of the as-built stainless steel shows a very fine dimples structure with 1.5μm in diameter. There was substantial ductility in all three specimens and no indication of brittle behavior. The shape of the stress-strain curves also indicates a ductile material with no brittle behavior in the joint in any stage.

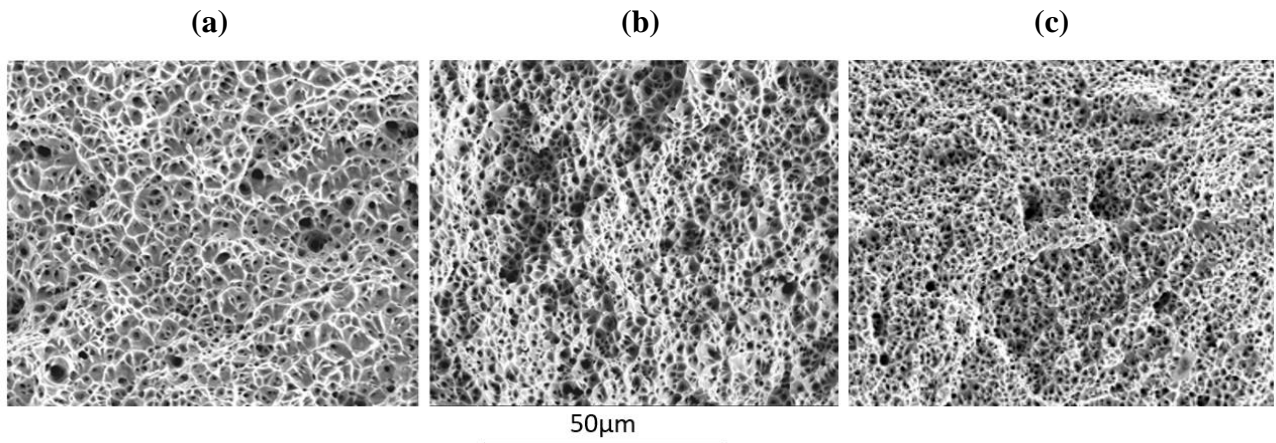

**Figure 14: Fractography of the (a) as-built interface, (b) aged sample, and (c) as-built stainless-steel base metal.**

Fracture surface EDS was performed in order to bring further insight into the difference in mechanical properties between the as-built and aged interface. Evidence for carbides were found in the dimples of broken tensile test aged sample. Elements indicating a mixture of both alloys were found. 12% Cr, 11% Ni and 1% Mn indicate the presence of stainless steel, while Mo and Co are components of the C300 maraging steelThe EDS analysis in Figure 15 (a) shows small amounts of 0.1%Ti.The concentration of Ti in the maraging steel is 1.1%. One may assume that the Ti also formed carbides as TiC. as Figure 15 (b) may suggest.

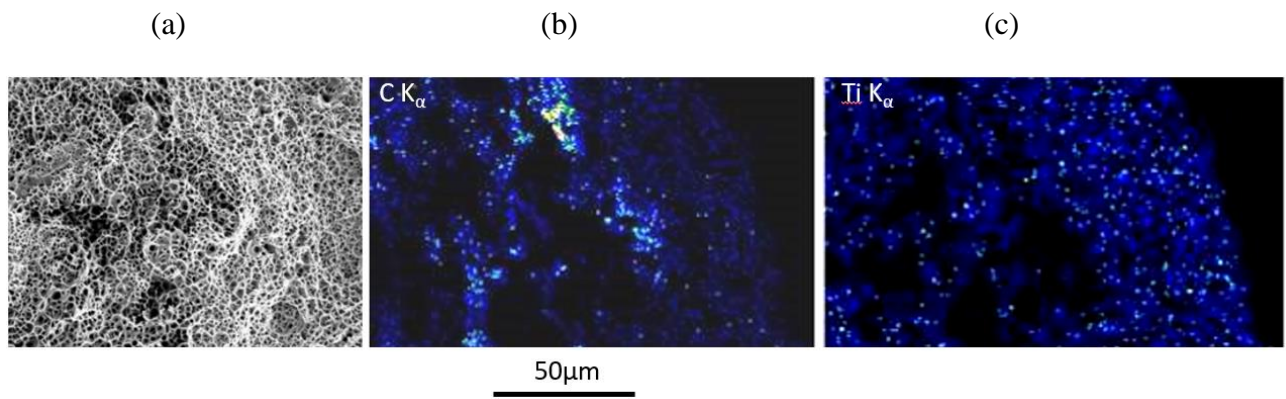

Figure 15: EDS scan maps of carbon and titanium distribution at aged interface (maraging steel side).

## 4. Discussion

This paper describes the development of additive manufacturing joint of C300 maraging steel to 316L austenitic stainless-steel using graded direct laser deposition technique. The development of the gradients were guided by thermodynamic calculations and the Schaeffler diagram. The presence of Ti in the interface zone might enhance the creation of Fe-Ti or $Fe_2Ti$, hard and brittle phases Ni-Ti or $Ti_2Ni$ or $TiNi_3$, Cr-Ti or $TiCr_2$ intermetallic. The gradient path was chosen based on the instrument functionality (two hoppers), powder availability, and brittle phase formation of the gradients. Despite the fact that the gradient passes through a $TiCr_2$ phase it was found that it does not lead to a fracture or failure during the print. In fact, it could be the case that the rapid cooling suppressed the formation of the intermetallic phase. According to the Schaeffler diagram SS316L is on the austenitic side. Depending on the exact composition of SS316, some percentage of ferrite can remain, but this was not observed. The differences between the thermodynamic calculations and the observed phases may be based on the fact that AM does not allow a full thermodynamic equilibrium due to fast cooling rates, and therefore not all phases predicted will form.

Despite no acute incompatibilities between the two alloys, characterization revealed that the way in which properties and microstructure vary along the composition gradient is neither linear nor trivial.

5. **Conclusions**

   5.1. Regardless of the individual composition of the layers deposited in the linear gradient, EDS measurements revealed that the actual composition gradient is almost continuous as a result of re-melting between adjacent layers.

   5.2. Phase distribution in the maraging steel and dual-phase regions are a consequence of the solidification substructure of the material resulting from solute redistribution of Ti, Mo, and Ni, limiting grain size to the scale of individual solidification cells.

   5.3. The tensile strength of the C300 to SS316L joint was found to be approximately equal to the as-built stainless steel while the decrease in elongation to fracture is due to the slight increase of Ti potentially creating carbides at interface and its vicinity.

   5.4. Solution treatment at 815°C had no effect on the gradient region's hardness or microstructure, which indicates that age hardening precipitates didn't form during graded AM deposition. The Austenizing of the C300 was not sufficient to dissolve unwanted Ti-rich nonmetallic inclusions such as Ti(C, N) and $Ti_4C_2S_2$.

   5.5. Age-hardening of the gradient joint couple at 482°C achieved hardness and strength values comparable to the conventionally C300 alloy, even though retained and reverted austenite were also observed. The age-hardening treatment had no effect on microstructure and properties of the 316L side and interface.

   5.6. Maraging heat treatment (aging) of the gradient joint couple, slightly reduced both the tensile strength and elongation to fracture of the joint.

   5.7. Thermodynamic modeling confirms that the risk of brittle intermetallic phase formation ($\sigma$-phase, $\chi$-phase, etc.) along the AM built gradient is no greater at intermediate compositions than it is for either base alloy. The linear composition path between the two alloys appears to be a favorable option.

6. **Acknowledgement**